\documentstyle[twoside,fleqn,espcrc2]{article}

\newcommand{\AmS}{{\protect\the\textfont2
  A\kern-.1667em\lower.5ex\hbox{M}\kern-.125emS}}

\title{Miscellanies of $ K^0 - \bar{K}^0 $ mixing and 
$ B_K $ \thanks{Talk presented by Weonjong Lee.}
}

\author{
	Markus Klomfass
	\address{Dept. of Physics,
	Columbia Univ., New York, NY 10027; 
	Veilchenweg 24, 65201 Wiesbaden, Germany.}
	and
	Weonjong Lee 
	\address{IBM, T.J. Watson Research Center,
	P.O. Box 218, Yorktown Heights, NY 10598, U.S.A.}
\thanks{Research sponsored in part by the U.S. Department of Energy.}
}

\begin{document}
 \def\thepage{IBM-HET-95-1}
 \thispagestyle{myheadings}

\input epsf

\begin{abstract}
We have computed $ B_K $, using two different methods with
staggered fermions on a $ 16^3 \times 40 $ lattice at $ \beta = 5.7 $
with two dynamical flavors of a mass 0.01.
Using an improved wall source method, we have studied a series of
non-degenerate quark antiquark pairs and observed no effect on $ B_K $,
although effects were seen on the individual terms making up $ B_K $.
\end{abstract}

\maketitle

\section{INTRODUCTION}
The standard model which is believed to describe 
our world of hadrons and leptons contains a number of
fundamental parameters.
The knowledge of hadronic weak matrix elements is crucial
to determine these parameters from the experiments.
Lattice gauge theory has reached a point that it is capable of
calculating hadronic weak matrix elements.
Especially, the knowledge of  $ B_K $ which describes the neutral
$ K $ meson mixing is crucial to narrow the domain of $ V_{td} $
and the top quark mass, which are the fundamental parameters of 
the standard model.
We have computed $ B_K $ on a $ 16^3 \times 40 $ lattice 
at $ \beta = 5.7 $ ($ a^{-1} \simeq 2.0 $ GeV)
with two dynamical flavors of a mass 0.01.
The results were obtained
over 155 gauge configurations.
Our work extends earlier calculations of $ B_K $ and 
includes experiments with alternative lattice formulations
and improved wall source methods.
There are two methods to transcribe
the continuum weak matrix elements
to the lattice with staggered fermions\cite{wlee0,sharpe0}: 
the one spin trace formalism
and the two spin trace formalism.
The two spin trace formalism (2TR)
has been used predominantly for weak matrix
element calculations\cite{sharpe1}.
Recently, the one spin trace formalism (1TR)
has been developed to a level which allows it to be used
for weak matrix element calculation on the lattice
\cite{wlee0}.
We have tried both formalisms to calculate $ B_K $.
The results are compared in this paper. 
We have studied an improved wall source named 
{\em cubic wall source} in order to project out a
specific hadronic state exclusively\cite{japan}.
The results are compared with those of the conventional
even-odd wall source.
We have looked into the effect of non-degenerate quark antiquark
pairs on $ B_K $ and the individual components making up $ B_K $
in detail.
From the standpoint of chiral perturbation theory,
the effects of non-degenerate
valence quark are related to the $ \eta' $ hairpin diagram in
(partially) quenched QCD\cite{sharpe2}.
We interpret our numerical results in terms of chiral
perturbation theory.
Preliminary results have already appeared in Ref.
\cite{wlee1}.
\section{$ K-\bar{K} $ Mixing}
In the continuum, $ B_K $ is defined as
\begin{eqnarray*}
B_K \equiv \frac{
\langle \bar{K}^0 \mid \bar{s} \gamma_\mu (1-\gamma_5) d \  \ 
\bar{s} \gamma_\mu (1-\gamma_5) d \mid K^0 \rangle }
{ \frac{8}{3}
\langle \bar{K}^0 \mid \bar{s} \gamma_\mu \gamma_5 d \mid 0 \rangle
\langle 0 \mid \bar{s} \gamma_\mu \gamma_5 d  \mid K^0 \rangle }
\end{eqnarray*}
The operator transcription of the continuum
$ \Delta S = 2 $ operator to the lattice in both one
spin trace formalism and two spin trace formalism
is explained in Ref. \cite{wlee0}.
\begin{figure}[t]
\vspace{-8mm}
\epsfxsize=70mm\epsfbox{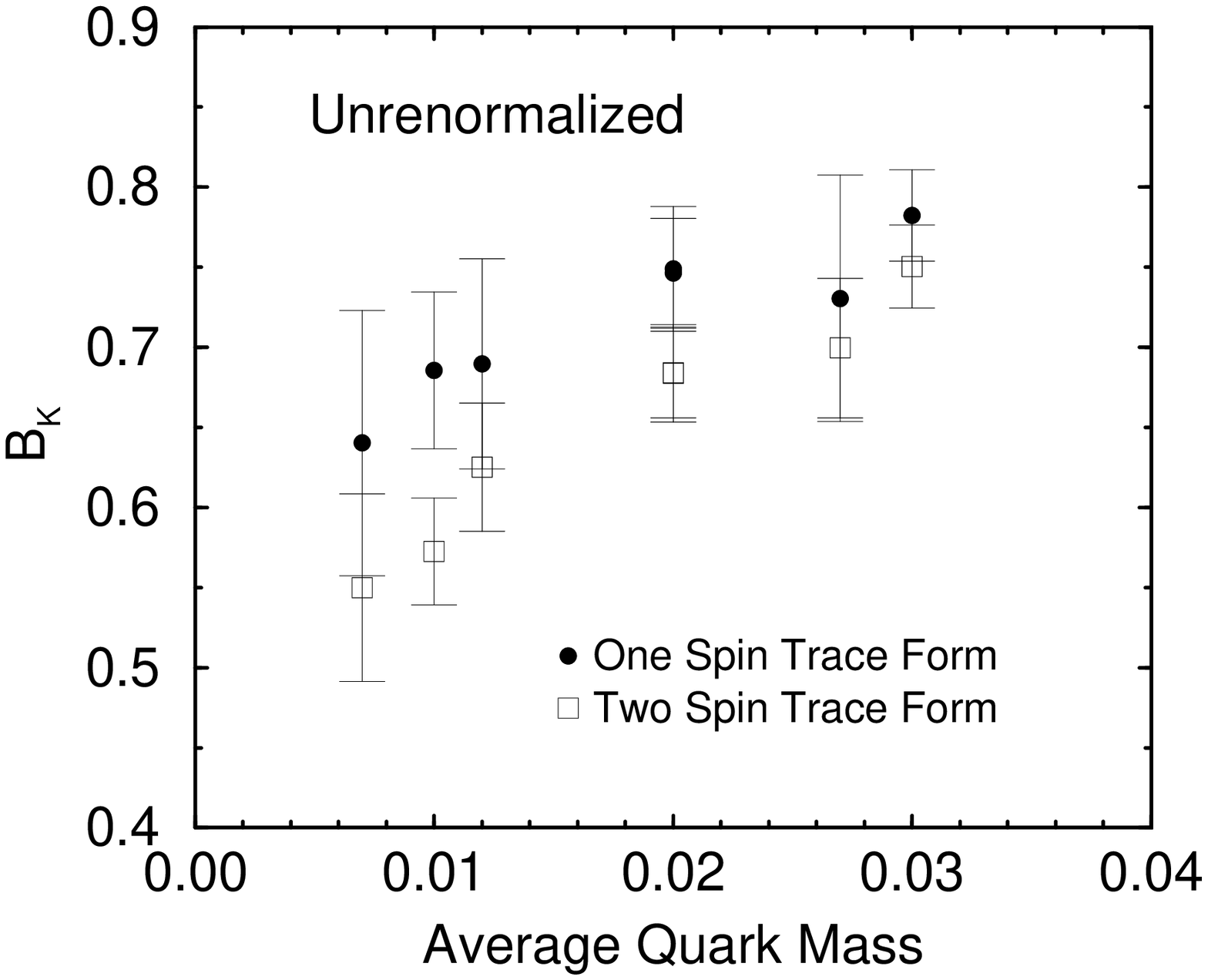}
\vspace{-3mm}
\epsfxsize=70mm\epsfbox{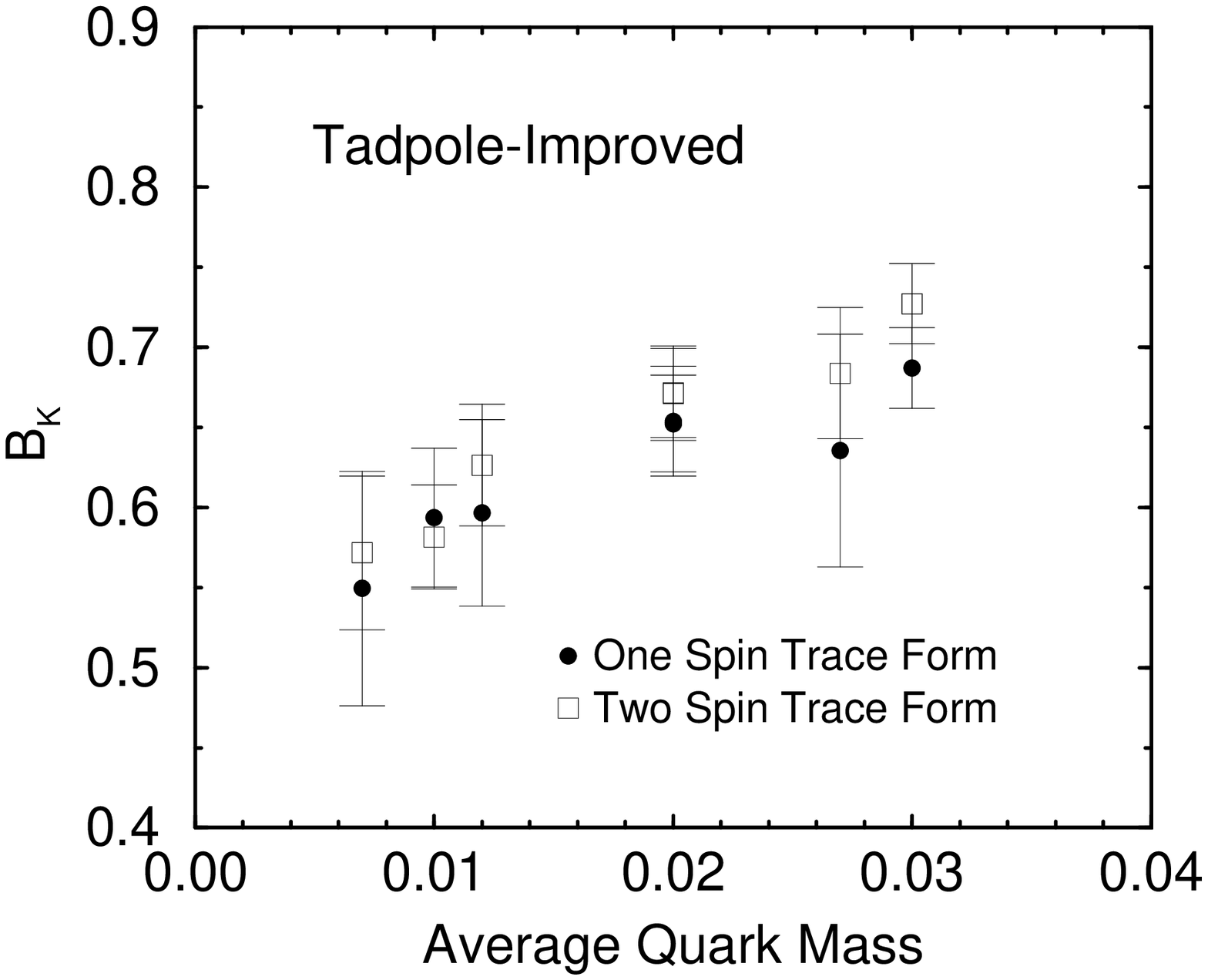}
\vspace{-10mm}
\caption{
Comparison of $ B_K $ in one spin trace form (filled circle)
with $ B_K $ in two spin trace form (empty square).
{\bf Top}: unrenormalized $ B_K $.
{\bf Bottom}: tadpole-improved renormalized $ B_K $
at $ \pi / a $ scale.
Even-odd wall source is used. 
}
\label{fig-1TR-2TR}
\end{figure}
The numerical results of $ B_K $ calculated in both formalisms
are compared in Figure \ref{fig-1TR-2TR}. 
The details of renormalization procedure is given in Ref. \cite{wlee0}.
From Figure \ref{fig-1TR-2TR},
note that numerical results of 
the tadpole-improved renormalized $ B_K $ in both formalisms
agree with each other better than those of unrenormalized $ B_K $. 
Here, we address two important questions
on the validity of our approach to $ B_K $.
  \pagenumbering{arabic}
  \addtocounter{page}{1}
The higher loop radiative correction of four fermion operators
cause the mixing of operators with different spin and flavor
structures.
The first question is how much the mixing of operators with different
flavor structure contributes to our weak matrix measurement
at finite lattice spacing non-perturbatively.
The second question is how exclusively we can select the
pseudo-Goldstone mode using our improved wall source
named {\em cubic wall source}.
We have chosen $ ( ( V+A ) \otimes S )^{\rm 2TR} $ in order to
address the above two questions, where we follow the notation
in Ref. \cite{wlee0}.
The matrix element of this operator with $ K $ mesons is supposed
to vanish in the continuum limit ($ a \rightarrow 0 $)
of lattice QCD, due to vanishing flavor trace.
\begin{figure}[t]
\vspace{-8mm}
\epsfxsize=70mm\epsfbox{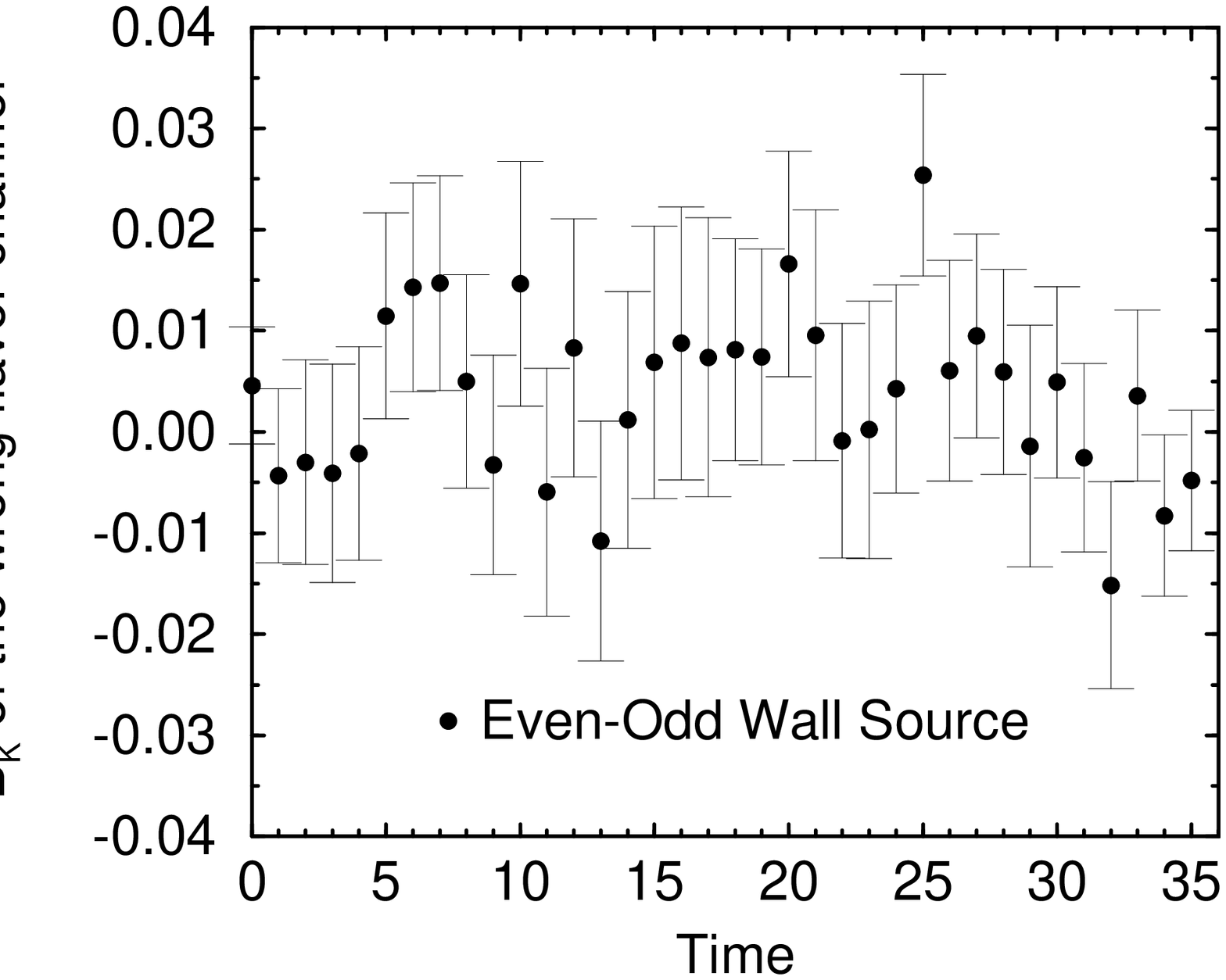}
\vspace{-3mm}
\epsfxsize=70mm\epsfbox{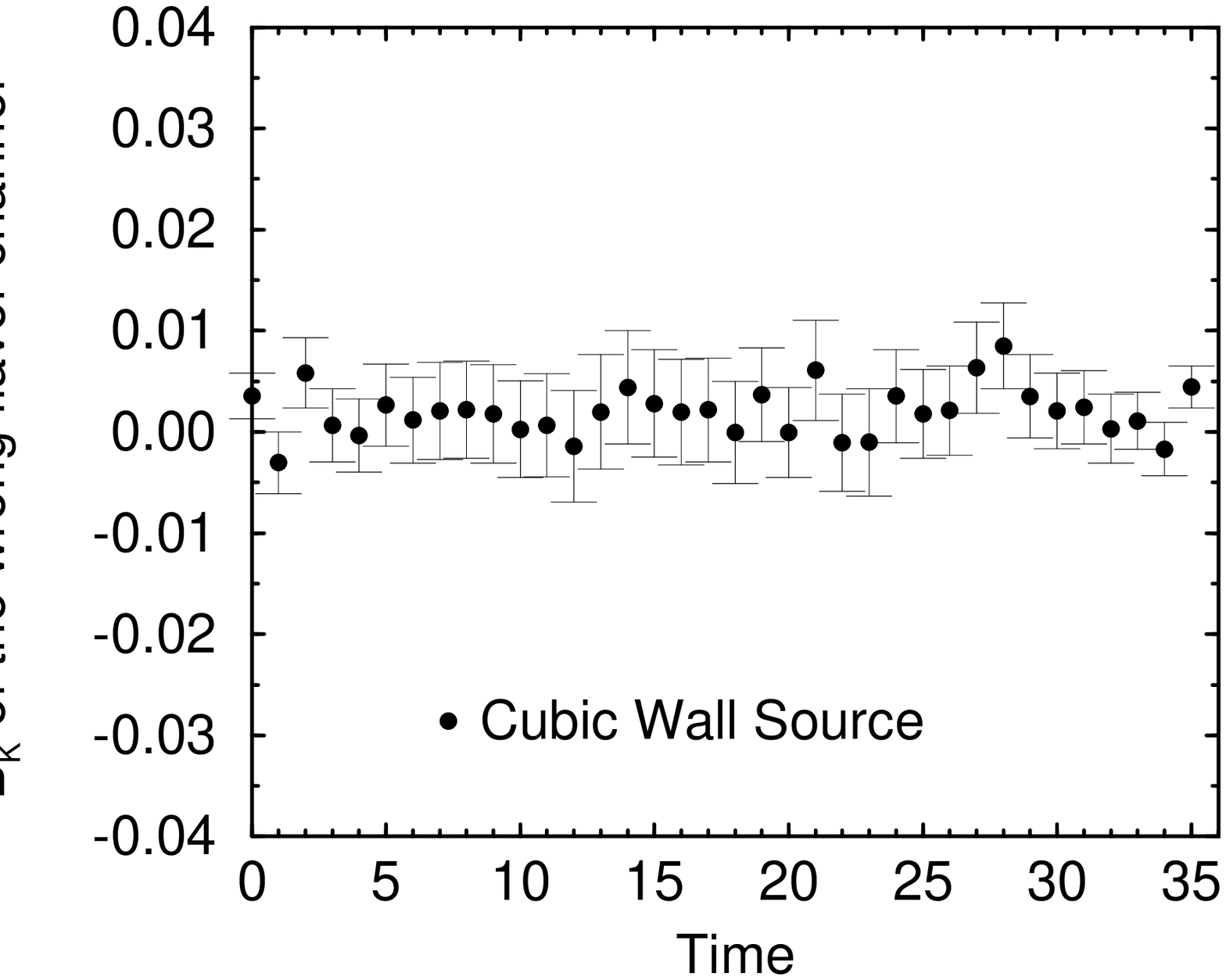}
\vspace{-10mm}
\caption{
Unrenormalized $ B_K $ of wrong (scalar-like)
flavor structure $ ((V+A) \otimes S)^{\rm 2TR} $
and of a valence quark mass 0.02
versus the Euclidean time.
Calculated in two spin trace form.
{\bf Top}: even-odd wall source is used.
{\bf Bottom}: cubic wall source is used. 
}
\label{source}
\end{figure}
We present the numerical results
of the wrong flavor channel in Figure \ref{source},
which tells us that 
the wrong flavor channel is highly suppressed
(within 1\% of $ B_K $) when cubic source is used.
This implies that unwanted mixing of
$ ((V+A) \otimes S)^{\rm 2TR} $ should be at most
1\% of $ B_K $ since it is suppressed by
$ \alpha_s/(4\pi) $ as well as by vanishing flavor
trace. 
The small deviation of the wrong flavor channel
from zero in Figure \ref{source} tells us how efficiently 
the improved wall sources suppress contaminations 
from unwanted hadronic states. 
\begin{figure}[t]
\vspace{-8mm}
\epsfxsize=70mm\epsfbox{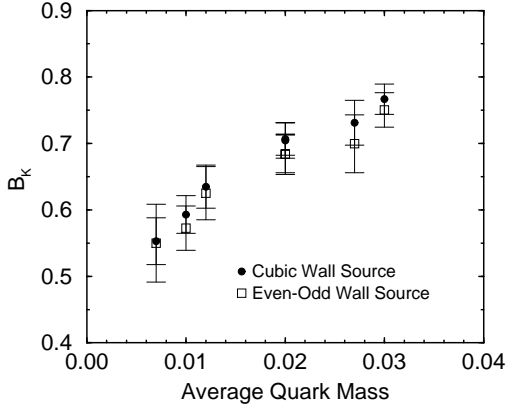}
\vspace{-10mm}
\caption{Unrenormalized $ B_K $ vs. average quark mass ($ m_q a $):
The filled circle (empty square) represents the data of 
cubic wall source (even-odd wall source). Calculated in the two
spin trace formalism.}
\label{bk-cub-ks}
\end{figure}
In Figure \ref{bk-cub-ks}, we compare 
$ B_K $ of cubic wall source 
with $ B_K $ of even-odd wall source.
Let us jump into the next issue: the effects of non-degenerate
quark antiquark pairs on $ B_K $ and its individual components,
and their relationship with various chiral logarithms.
In our numerical simulation, we have used three degenerate
quark antiquark pairs: \{0.01, 0.02, 0.03\}, and 
four non-degenerate pairs: \{(0.004, 0.01), (0.04,0.02),
(0.01, 0.03), (0.01,0.03)\} to produce $ K $ meson state.
In order to discuss individual components of $ B_K $
in an organized way, we need to consider a theory with four valence
flavors\cite{sharpe3}: $ S $ and $ S' $ both with mass $ m_{s} $
as well as $ D $ and $ D' $ both with mass $ m_{d} $.
Let $ K^0 $ be the $ \bar{S} \gamma_5 D $ pion and
$ K'^0 $ be the corresponding state with primed quarks
($ \bar{S}' \gamma_5 D' $).
We define individual components of $ B_K $ as
\begin{eqnarray*}
{\cal V.S.} & \equiv & \hspace*{-3mm} \frac{4}{3}
\langle \bar{K}'^0 \mid \bar{S}'_a \gamma_{\mu} \gamma_5 D'_a \mid 0 \rangle
\langle 0 \mid \bar{S}_b \gamma_{\mu} \gamma_5 D_b \mid K^0 \rangle
\\
B_{V1} & \equiv & \langle \bar{K}'^0 \mid \bar{S}'_a \gamma_\mu  D'_b
\ \ \bar{S}_b \gamma_\mu D_a \mid K^0 \rangle / {\cal V.S.}
\\
B_{V2} & \equiv & \langle \bar{K}'^0 \mid \bar{S}'_a \gamma_\mu  D'_a
\ \ \bar{S}_b \gamma_\mu D_b \mid K^0 \rangle / {\cal V.S.}
\\
B_{A1} & \equiv & 
\langle \bar{K}'^0 \mid \bar{S}'_a \gamma_\mu \gamma_5 D'_b
\ \ \bar{S}_b \gamma_\mu \gamma_5 D_a \mid K^0 \rangle / {\cal V.S.}
\\
B_{A2} & \equiv & 
\langle \bar{K}'^0 \mid \bar{S}'_a \gamma_\mu \gamma_5 D'_a
\ \ \bar{S}_b \gamma_\mu \gamma_5 D_b \mid K^0 \rangle / {\cal V.S.}
\\
B_V & = &  B_{V1} + B_{V2} \\
B_A & = &  B_{A1} + B_{A2} \\
B_K & = &  B_V + B_A
\end{eqnarray*}
where $ a, b $ represent color indices.
\begin{figure}[t]
\vspace{-8mm}
\epsfxsize=70mm\epsfbox{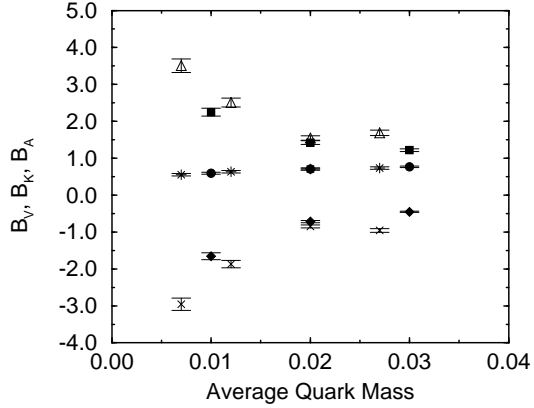}
\vspace{-10mm}
\caption{
Unrenormalized $ B_K $, $ B_V $, $ B_A $ vs. average quark mass:
Calculated using cubic wall source in the two spin trace formalism.
$ B_A $: empty triangle and filled square.
$ B_K $: star and filled circle.
$ B_V $: x and filled diamond.
Filled square, circle, and diamond represent degenerate
quark antiquark pairs. All other symbols correspond to
non-degenerate pairs. 
}
\label{bk-bv-ba}
\end{figure}
Let us summarize chiral perturbation results for the above
individual components,
the details of which are explained in Ref. \cite{sharpe3}.
In full QCD chiral perturbation, 
the one-loop corrections to $ B_{V1} $, $ B_{V2} $,
$ B_{A1} $, and $ B_{A2} $ include a logarithmically divergent term 
in the chiral limit $ m_K \rightarrow 0 $, which is called an
{\em enhanced chiral logarithms}, and which is absent in $ B_K $.
The enhanced chiral logarithms are not a function of quark mass difference
between $ s $ and $ d $ quarks.
We present $ B_K $, $ B_V $, and $ B_A $ data
versus quark mass in Figure \ref{bk-bv-ba}, which 
illustrates the existence of a divergence in $ B_V $ and $ B_A $
and no divergence in $ B_K $ in the chiral limit.
There are two possible differences between partially quenched or
quenched and full QCD.
The first difference is that the meson spectrum eigenstates are not
same.
The second difference comes from $ \eta' $ loops, which can not
contribute in full QCD because it is too heavy for chiral dynamics.
The hairpin diagram contribution to $ B_K $ and its individual component,
which is called ``{\em (partially) quenched chiral log}",
vanishes in the limit of $ m_s = m_d $.
In other words, (partially) quenched chiral logarithms are functions
of quark mass difference.
Since chiral perturbation theory
predicts that $ B_{A1} $ has a constant term about 1/3 smaller
and a enhanced chiral logarithmic term around 3 times larger than
$ B_{A2} $, we have chosen $ B_{A1} $ as a useful measurement
adequate to observe, if present,
both enhanced and quenched chiral logarithms.
\begin{figure}[t]
\vspace{-8mm}
\epsfxsize=65mm\epsfbox{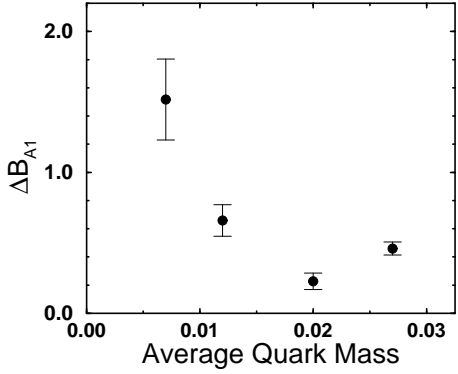}
\vspace{-10mm}
\caption{$ \Delta B_{A1} $ vs. average quark mass:
Calculated using cubic wall source
in the two spin trace formalism.}
\label{DBA1}
\end{figure}
In order to observe the effect of non-degenerate quark antiquark
pairs on $ B_{A1} $, first we fit data of degenerate pairs to
obtain $ B^{\rm deg}_{A1} (m_K) $ and second we subtract 
those degenerate contributions from non-degenerate data as follows:
\begin{eqnarray*}
\Delta B_{A1} (m_K, \epsilon) & \equiv & 
\frac{ B_{A1}(m_K, \epsilon) - B^{\rm deg}_{A1} (m_K) }
	{ \epsilon },
\end{eqnarray*}
where $ B_{A1}(m_K, \epsilon) $ is data of non-degenerate quark
antiquark pairs and $ \epsilon \equiv (m_s - m_d)/(m_s + m_d) $.
We present $ \Delta B_{A1} $ data in Figure \ref{DBA1},
which indicates that there is an additional divergence as
a function of $ \epsilon $.
This additional divergence might come from (partially)
quenched chiral logarithms or from a finite volume effect.
This needs more numerical evidence and further 
theoretical understanding.
In order to detect the effect of non-degenerate quark antiquarks
on $ B_K $, we follow the procedure similar to $ B_{A1} $ case.
First, we fit degenerate date to $ B^{\rm deg}_K (m_K) $.
Second, we introduce $ \Delta B_K $ as follows:
\begin{eqnarray*}
\Delta B_K (m_K) \equiv
\frac{ B_K ( m_K, \epsilon) - B^{\rm deg}_K (m_K) }
	{ \epsilon^2 },
\end{eqnarray*}
where the function is normalized by $  \epsilon^2 $ since
chiral perturbation predicts that leading effect of non-degenerate
quark antiquark pairs on $ B_K $ is of order $ \epsilon^2 $.
\begin{figure}[t]
\vspace{-8mm}
\epsfxsize=65mm\epsfbox{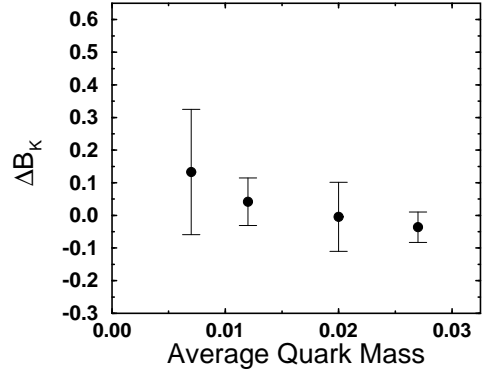}
\vspace{-10mm}
\caption{$ \Delta B_K $ versus average quark mass:
Calculated using cubic wall source in the two spin trace formalism.
}
\label{DBK}
\end{figure}
We present the $ \Delta B_K $ data in Figure \ref{DBK},
which illustrates that the non-degenerate quark mass
effect is much smaller than our statistical error.
Our best results are  
unrenormalized $ B_K (m_K) $ = 0.658(77)
and tadpole-improved renormalized (N.D.R.)
$ B_K( m_K, \mu=\frac{\pi}{a} ) $
= 0.659(63).
One of the authors (W. Lee) wants to express heartful
gratitude to Prof. N.H. Christ, R.D. Mawhinney,
D. Zhu, D. Chen, S. Chandrasukaharan,
and Z. Dong for their kind help.

\begin{thebibliography}{99}
%
%
%
\bibitem{wlee0} W. Lee and M. Klomfass, Phys. Rev. D {\bf 51} (1995)
6426.
%
\bibitem{sharpe0} S. Sharpe {\em et al.}, Nucl. Phys. {\bf B286}
(1987) 253.
%
\bibitem{sharpe1} S. Sharpe, Nucl. Phys. B (Proc. Suppl.) {\bf 34}
(1994) 403; N. Ishizuka {\em et al.}, Phys. Rev. Lett. {\bf 71}
(1993) 24; G. Kilcup, Phys. Rev. Lett. {\bf 71} (1993) 1677.
%
\bibitem{japan} M. Fukugita {\em et al.}, Phys. Rev. D {\bf 47}
(1993) 4739.
%
\bibitem{sharpe2} S. Sharpe, Phys. Rev. D {\bf 46}, (1992) 853. 
%
\bibitem{wlee1} W. Lee and M. Klomfass, Nucl. Phys. B (Proc. Suppl.)
{\bf 42} (1995) 418.
%
\bibitem{sharpe3} S. Sharpe, Phys. Rev. D {\bf 46}, (1992) 3146.
%
%
%
%
\end{thebibliography}
\end{document}